\documentclass[%
 reprint,
superscriptaddress,
nofootinbib,
 amsmath,amssymb,
 aps
]{revtex4-2}

\usepackage{graphicx}
\usepackage{dcolumn}
\usepackage{bm}
\graphicspath{{Figures/}}
\usepackage{siunitx}

\begin{document}

\preprint{APS/123-QED}

\title{Design of axion and axion dark matter searches based on ultra high Q SRF cavities}
\author{B. Giaccone}
\email{giaccone@fnal.gov}
\author{A. Berlin}
\author{I. Gonin}
\author{A. Grassellino}
\author{R. Harnik}
\affiliation{Fermi National Accelerator Laboratory, Batavia, IL 60510, USA}
\author{Y. Kahn}
\affiliation{University of Illinois Urbana-Champaign, Champaign, IL 61801, USA}
\author{T. Khabiboulline}
\author{A. Lunin}
\author{A. Melnychuk}
\author{A. Netepenko}
\author{R. Pilipenko}
\author{Y. Pischalnikov}
\author{S. Posen}
\author{O. Pronitchev}
\author{A. Romanenko}
\author{V. Yakovlev}
\affiliation{Fermi National Accelerator Laboratory, Batavia, IL 60510, USA}

\date{\today}

\begin{abstract}
The Superconducting Quantum Materials and Systems center is developing searches for dark photons, axions and ALPs with the goal of improving upon the current state-of-the-art sensitivity. These efforts leverage on Fermi National Accelerator expertise on ultra-high quality factor superconducting radio frequency cavities combined with the center research on quantum technology. Here we focus on multiple axion searches that utilize $\approx 10^{10}$ quality factor superconducting radio frequency cavities and their resonant modes to enhance the production and/or detection of axions in the cavity volume. In addition, we present preliminary results of single-mode and multi-mode nonlinearity measurements that were carried out as part of an experimental feasibility study to gain insight on the behavior of the ultra-high quality factor resonators and the experimental RF system in the regime relevant for axion searches.

\end{abstract}

\maketitle

\section{\label{sec:intro}Introduction}
Superconducting radio frequency (SRF) cavities are routinely used in particle accelerators, oprating at accelerating gradients of the order of tens of MV/m and intrinsic quality factors as high as $10^{11}$ \cite{padamseerf}. As part of the Superconducting Quantum Materials and Systems (SQMS) Physics and sensing thrust, we are now looking at applications of SRF cavities for searches of beyond the standard model particles \cite{berlin2022searches, posen2022measurement}. Searches for axions are based on the theoretical prediction that this hypothesized particle can couple to photons in presence of external magnetic field \cite{peccei1977constraints, peccei1977cp}. This paper will focus on multiple axion and axion dark matter searches \cite{berlin2020axion, bogorad2019probing, gao2021axion}, all based on ultra high quality factor ($Q_0$) SRF cavities. The searches here discussed use the resonant modes of the cavity to try to detect, and in some case produce, axions rather than applying an external magnetic field on the niobium cavity. Details of the searches are discussed in Sec. \ref{sec:searchdesign}.

In addition to the cavity's design, the preliminary results of an ongoing feasibility study are presented here. The scope of this study is to identify possible mechanisms of energy leakage between the cavity resonant modes and between excited modes and their linear combinations. If particular linear combinations of the pump modes are excited in the cavity, they could mimic an axion signal or they could greatly limit the search sensitivity. Sauls discusses this topic in \cite{sauls2022theory}. In addition to nonlinearities, we are also interested in studying possible noise sources that could limit the search reach by for example raising the background floor, exciting spurious frequencies in the cavity, causing crosstalk through the cavity or to the readout instrumentation.

\section{\label{sec:nonlinstudy}Feasibility study}
\begin{figure}
    \centering
    {\includegraphics[width=1\columnwidth]{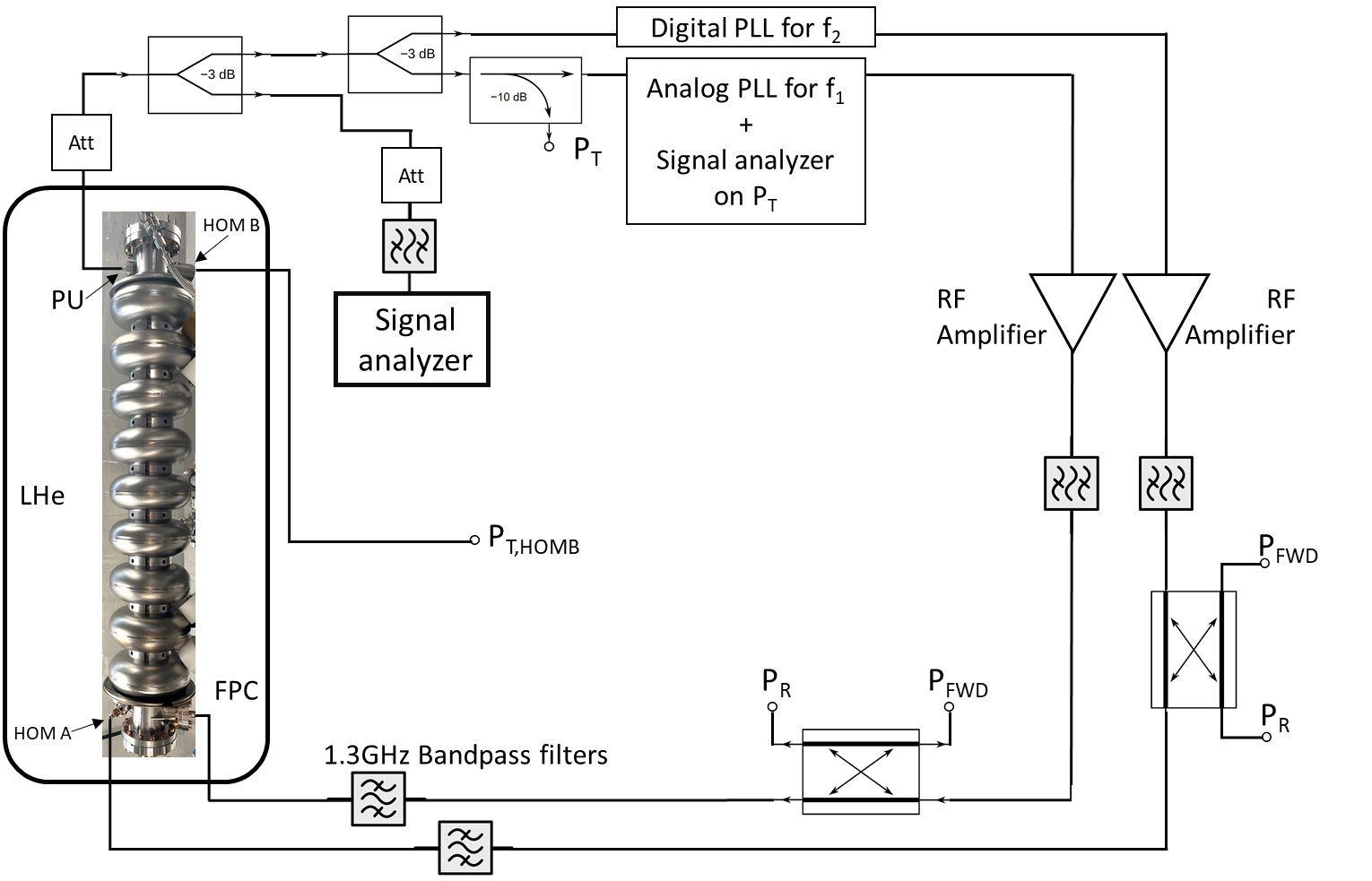}} \quad
	\caption{Simplified schematic of the experimental system used to conduct one-mode and two-mode measurements on the 9-cell cavities. The abbreviation PLL in the schematic is used to identify the phase-locked loop block used to track and lock on the cavity resonance frequency.}
	\label{fig:expsystem}
\end{figure}

As part of the preparatory study, measurements with one or two excited modes were conducted on 9-cell Nb \SI{1.3}{\giga\hertz} TESLA shape cavities \cite{aune2000superconducting}. The pump modes used for the measurements belong to the first monopole $\mathrm{{TM}_{010}}$ cavity passband: a family of nine transverse magnetic modes around \SI{1.3}{\giga\hertz} and with different relative phases between the nine cells of the cavity. We refer to these nine modes as $\frac{1}{9} \pi$, $\frac{2}{9} \pi$, $\frac{3}{9} \pi$, ..., up to $\pi$, where $\frac{1}{9} \pi$ is the $\mathrm{{TM}_{010}}$ with lowest frequency ($\approx \SI{1.27}{\giga\hertz}$), while the $\pi$ mode is the $\mathrm{{TM}_{010}}$ with highest frequency ($\approx \SI{1.30}{\giga\hertz}$) and with $180 ^{\circ}$ phase difference between neighboring cells. The tests were conducted at \SI{2}{\kelvin} in Fermi National Accelerator Laboratory (FNAL) Vertical Test Stand (VTS) facility. The experimental setup used for single-mode and multi-mode measurements is shown in Fig. \ref{fig:expsystem}.

\subsection{Single-mode Measurements}
\begin{figure}
    \centering
    {\includegraphics[width=0.6\columnwidth]{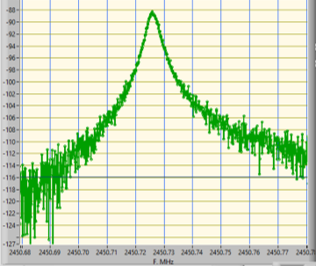}} \quad
	\caption{Example of measurement conducted at \SI{2}{\kelvin} of one of the cavity resonant modes that will be used in the single-mode measurements.}
	\label{fig:2450mhzmeas}
\end{figure}

The scope of the one-mode measurements is to study possible energy transfer between one excited mode and other resonant modes of the cavity. In preparation for the searches proposed in \cite{berlin2020axion, bogorad2019probing, gao2021axion}, we are interested in studying energy transfer to modes with $\Delta f \approx 1-\SI{10}{\mega\hertz}$ for Berlin \textit{et al.} search \cite{berlin2020axion} and $ \Delta f \approx \SI{1}{\giga\hertz}$ for Gao and Harnik search \cite{gao2021axion}, where $\Delta f = |f_{pump} - f_{2}|$. Since this experiment is being conducted on \SI{1.3}{\giga\hertz} 9-cell cavities, we have access to nine modes in the $\mathrm{{TM}_{010}}$, spread across $\approx \SI{30}{\mega\hertz}$ from \SI{1.27}{\giga\hertz} to \SI{1.3}{\giga\hertz}. To study the range $ \Delta f \approx \SI{1}{\giga\hertz}$ we identified a series of modes with resonant frequencies around $\SI{2.4}{\giga\hertz}$. Fig. \ref{fig:2450mhzmeas} shows one of the resonant modes ($f_{res}=\SI{2450.72}{\mega\hertz}$ at \SI{2}{\kelvin}) studied during the one-mode measurements.

Measurements in the range $\Delta f \approx 1-\SI{10}{\mega\hertz}$ were limited in sensitivity by the fact that the noise of the amplifier used to amplify the RF signal on the cavity input was sufficient to excite all nine $\mathrm{{TM}_{010}}$ modes. Due to the proximity of these peaks, they all fall within the range of operation of the RF amplifiers at our disposal. Multiple RF amplifiers were tested for this measurements, obtaining similar results of exciting all nine $\mathrm{{TM}_{010}}$ modes at different levels. When using an RF source directly connected to the cavity input, without intermediate amplification, we were able to prevent excitation of the $\mathrm{{TM}_{010}}$ modes other than the pump mode. In this case the input power was limited to \SI{880}{\micro\watt}, with $P_{cavity} \approx \SI{500}{\micro\watt} = -3\,$dBm. When pumping the $\pi$ mode using only the RF source, we successfully measured no excitation above background of the other eight $\mathrm{{TM}_{010}}$ resonant modes, setting a limit on the energy transfer between an excited mode and nearby (from 1 to \SI{30}{\mega\hertz} apart) resonant modes. However, when the $\pi$ mode is excited, we find that the background throughout the range of the \SI{1.3}{\giga\hertz} passband is raised, moving the background 10 to 20 dB above the level of the expected room temperature background expected from a \SI{1}{\hertz} bandwidth measurement. This translates in a corresponding reduction of our limit, setting the ratio of pump power to signal power to $\approx 150\,$dB.

\begin{figure*}
    \centering
    {\includegraphics[width=1\textwidth]{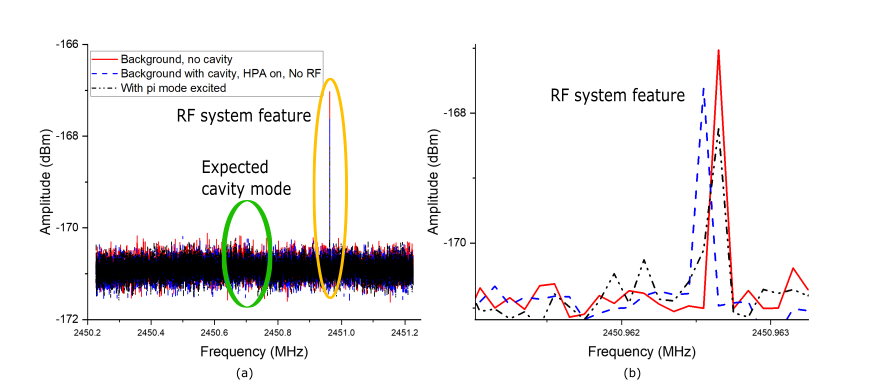}} \quad
	\caption{In panel (a) are shown the results of the single-mode measurements in the frequency range of $2450-$\SI{2451}{\mega\hertz}. No excitation of the resonant mode at \SI{2450.7}{\mega\hertz} is measured above the room temperature background level, allowing to set a limit on the possible energy transfer from a pump mode to other cavity resonant modes $\approx \SI{1}{\giga\hertz}$ away. A chain of RF filters was placed between the cavity and the SA in order to filter out the \SI{1.3}{\giga\hertz} pump mode from the SA input line. The entire RF filter chain was measured to have $\approx 5dB$ insertion loss in the $\SI{2.4}{\giga\hertz}$ frequency range. This translates into an equivalent sensitivity decrease on the energy leakage that could be measured from the pump mode to the $\SI{2.4}{\giga\hertz}$ resonant modes. Panel (b) shows a zoom of the area highlighted in yellow: the peaks measured around \SI{2450.9}{\mega\hertz} are found to be a feature of the RF system, generated outside of the cavity, as indicated by the fact that these peaks are measured also when no RF is sent in input to the cavity and when the cavity is not connected to the read-out instrument (signal analyzer). For both measurements a bandwidth value of 1Hz was used for the SA.}
	\label{fig:1modemeas}
\end{figure*}

As for the range $ \Delta f \approx \SI{1}{\giga\hertz}$, we studied possible energy leak from the pump mode ($\pi$ mode, ninth mode of the $\mathrm{{TM}_{010}}$ band) to a few resonant modes around $\SI{2.4}{\giga\hertz}$. The $\pi$ mode (with frequency $\approx$ \SI{1.3}{\giga\hertz}) was used as pump mode for these measurements with input power $P_{in} \approx \SI{8}{\watt}$ and $Q_0 = 2 \times 10^{10}$. An example of the results of these measurements is shown in Fig. \ref{fig:1modemeas}. For all the modes studied we did not measure any excitation of the peak above background level, setting a limit on the possible energy leakage from the pump mode to other resonant modes. However, the limit is not presented in this work since it is still preliminary and further measurements are needed to confirm it. The peaks highlighted in Fig. \ref{fig:1modemeas} were found to be transmitted to the signal analyzer (SA) outside of the cavity, likely through the RF line that connects the cavity to the VTS RF system. This was verified by disconnecting the cavity from the signal analyzer and repeating the measurement. In this configuration we found that we could still measure the peaks in the SA, confirming that they were not being transmitted to the SA through the cavity.

\subsection{Multi-mode Measurements}
The two-mode measurements are relevant for Bogorad \textit{et al.} \cite{bogorad2019probing} proposed search, as the scope is to measure excitation of linear combinations of the pump modes. For these measurements, two different antennas and input lines are used to excite the two pump modes in the cavity. One mode is excited through the fundamental power coupler (FPC), the other mode is excited through one higher order mode (HOM) antenna. The FPC is the antenna routinely used as input coupler in SRF cavities, its length is chosen such that the external quality factor of the antenna is almost critically matched with the internal cavity quality factor for the $\mathrm{{TM}_{010}}$ at \SI{2}{\kelvin}. The HOM antenna instead is used in accelerators to extract power from all modes except the $\pi$. While the $\pi$ mode is excited in the cavity volume and used to accelerate the particle beam, spurious modes can be excited by passage of the beam and if the power from these modes is not extracted, particles from beam can interact with the higher order modes electromagnetic fields, degrading the beam quality. HOM antennas are designed to offer good coupling across a vast range of frequency in order to extract power from these excited spurious modes. In addition, the HOM antennas have a notch filter that is tuned to the $\pi$ frequency, allowing to reject the extraction of power from the accelerating mode.

The pump modes used for the 2-mode measurements were the $\pi$ mode on the FPC and the $\frac{4}{9} \pi$ or the $\frac{5}{9} \pi$ on the HOM coupler. The HOM antenna was tuned so that its notch filter would reject the $\pi$ mode and preventing leakage of power from the FPC to the HOM coupler. The notch filter has a finite bandwidth that prevented us from using modes close to the $\pi$ as second input mode, hence the choice of the $\frac{4}{9} \pi$ and the $\frac{5}{9} \pi$. With both modes excited in the cavity, we investigated the excitation of linear combination of the pump modes. These linear combinations in general do not coincide with resonant modes of the cavity. However, in the case of $2\omega_1 - \omega_2$ and $2\omega_2 - \omega_1$ they fall within few MHz from the pump modes and as such it is nontrivial to attenuate the pump modes from the input of the SA, still allowing the target linear combinations to be transmitted to the SA. Failure to attenuate the pump modes sufficiently on the SA input line results in the intermodulation of the pump modes in the SA itself, which generates peaks at the linear combinations. Our effort to attenuate the pump modes was not sufficient to prevent the SA intermodulation. As a result we could not set a limit on the energy transfer between pump modes and their linear combinations in the nearby frequency range.

\begin{figure}[!]
    \centering
    {\includegraphics[width=1\columnwidth]{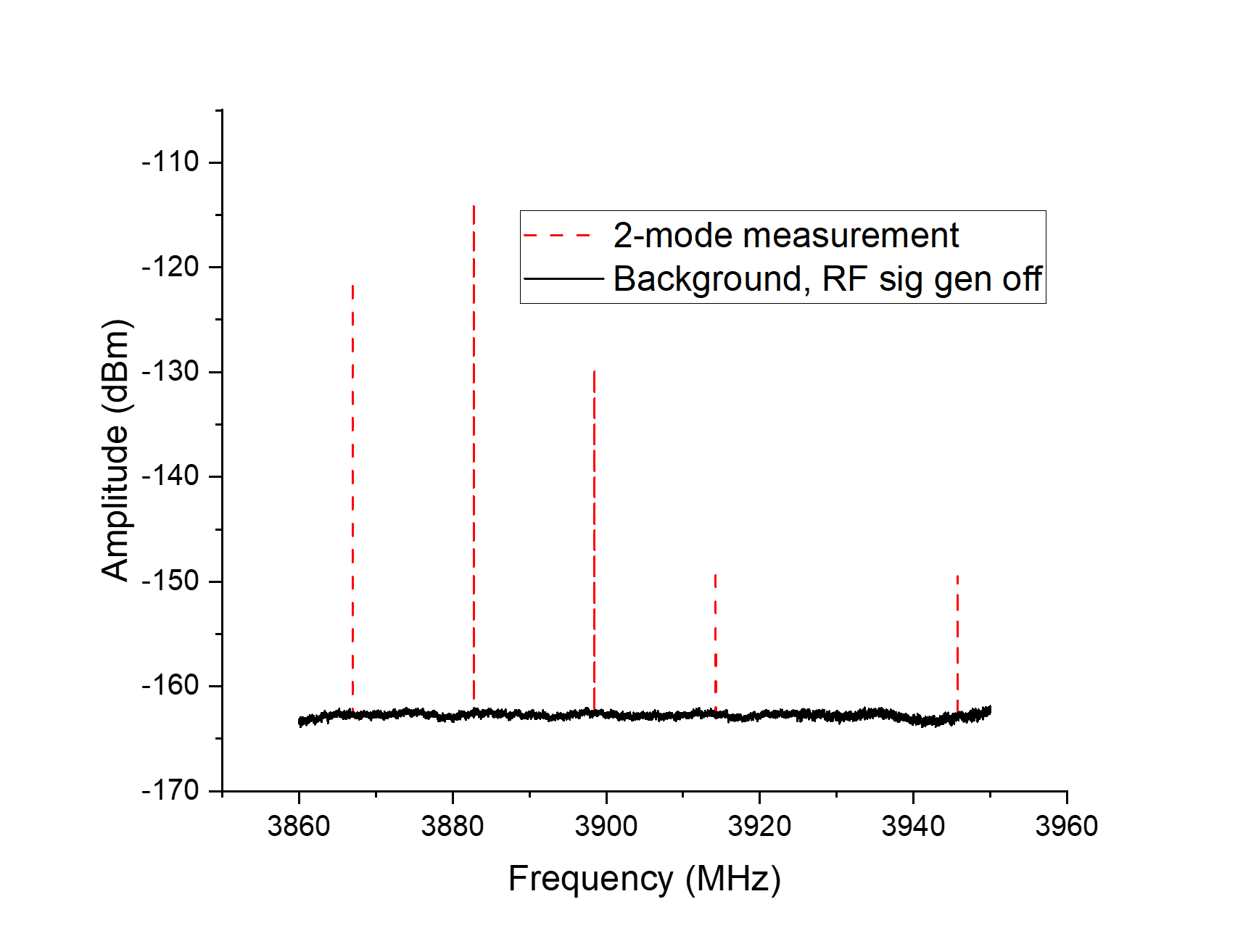}} \quad
	\caption{Example of two-mode measurement power spectrum measured in the \SI{3.9}{\giga\hertz} range, where the $2\omega_1 + \omega_2$ and $2\omega_2 + \omega_1$ are expected. In addition to these two combinations, we also measure $3\omega_1$, $3\omega_2$ and higher orders as for example $4*\omega_1 - \omega_2$. Here a bandwidth (BW) value of 1Hz was used for the SA, hence the higher background level compared to Fig. \ref{fig:1modemeas}. Measurements with BW=1Hz were also conducted on smaller frequency ranges centered on individual linear combinations.}
	\label{fig:2modemeas}
\end{figure}

We also investigated the excitation of linear combinations at $ \Delta f \approx \SI{2.6}{\giga\hertz}$ away from the pump modes, by studying the $2\omega_1 + \omega_2$ and $2\omega_2 + \omega_1$ range. In this case it was possible to suppress the pump modes from the cavity output by $\approx$ 120\,dB. Fig. \ref{fig:2modemeas} shows an example of results of the two-mode measurements in the \SI{3.9}{\giga\hertz} range and compares the signal measured on the SA when the RF source is off and no modes are excited in the cavity, with the power spectrum measured when two modes are pumped in the cavity volume. In this case $\pi$ and $\frac{4}{9} \pi$ are used as pump modes, both with $P_{in}=\SI{4}{\watt}$. Once again the SA measured linear combinations of the pump modes. It is not straightforward to identify the origin of these peaks and this matter is still under study. The amplitude of the linear combinations around \SI{3.9}{\giga\hertz} is in some cases higher than the amplitude of the pump modes that the SA receives on the input line, suggesting that SA intermodulation can not be the only source generating these linear combinations. An initial study of crosstalk between input lines and SA seems to exclude this as possible source, however, we have not yet ruled out crosstalk between the two input lines or crosstalk between input lines and cavity output. We also conducted an initial series of decay measurements where we turned off the input power in one of the pump modes and measured the decay time of a linear combination. However, this study was not conclusive since the decay time of the pump mode in the cavity would reflect on the decay time of the linear combination even if this peak was generated after the cavity (for example in the SA or due to crosstalk between cavity output and cavity input lines). A thorough study of the power dependency of the measured linear combinations as a function of pump modes power, combined with a more detailed series of decay measurements of the linear combinations, may possibly help in pinpointing their source.

Although this feasibility study was not conclusive yet in confirming or fully excluding sources of nonlinearities in the cavity, it was extremely valuable in guiding the design of the RF system that will be procured for the searches and it gave precious input on RF design of the cavities as it highlighted multiple sources of noise that could be transmitted to the cavity or to the readout equipment.

\section{\label{sec:searchdesign}Cavity RF and mechanical design}
The schemes proposed in \cite{berlin2020axion, bogorad2019probing, gao2021axion} require cavity design tailored to each search. Here we present designs for the searches in \cite{berlin2020axion} and \cite{bogorad2019probing}, while the design for \cite{gao2021axion} is planned to start in the near future.

\subsection{Two-Mode Axion Dark Matter Search}
\begin{figure}
    \centering
    {\includegraphics[width=0.7\columnwidth]{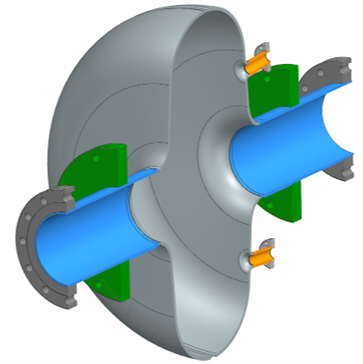}} \quad
	\caption{Finalized geometry of the single cell cavity designed for the two-mode axion dark matter search. Colored in grey is the cavity cell, in light blue the beam tubes, in dark grey the vacuum flanges, in green the flanges for the tuner and in orange the coupler ports for the loop antennas.}
	\label{fig:2modegeom}
\end{figure}

\begin{figure}
    \centering
    {\includegraphics[width=0.7\columnwidth]{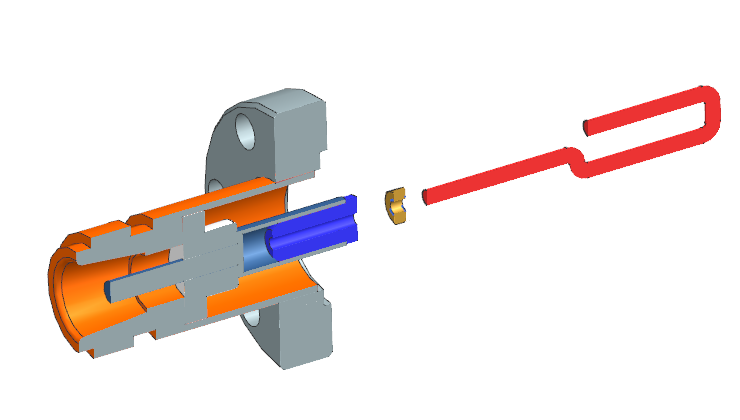}} \quad
	\caption{Exploded schematic of the loop antenna that will be used to couple to the $\mathrm{{TE}_{011}}$ magnetic field.}
	\label{fig:2modeantenna}
\end{figure}

The search proposed by Berlin \textit{et al.} in \cite{berlin2020axion} requires a cavity where two resonant modes, one a TE, one a TM, have almost degenerate frequency. Berlin \textit{et al.} search is based on the principle of the heterodyne detection \cite{berlin2021heterodyne} where the axion mass that the detector is sensitive to is determined by the frequency splitting between the pump and the signal mode.
\begin{figure}
    \centering
    {\includegraphics[width=1\columnwidth]{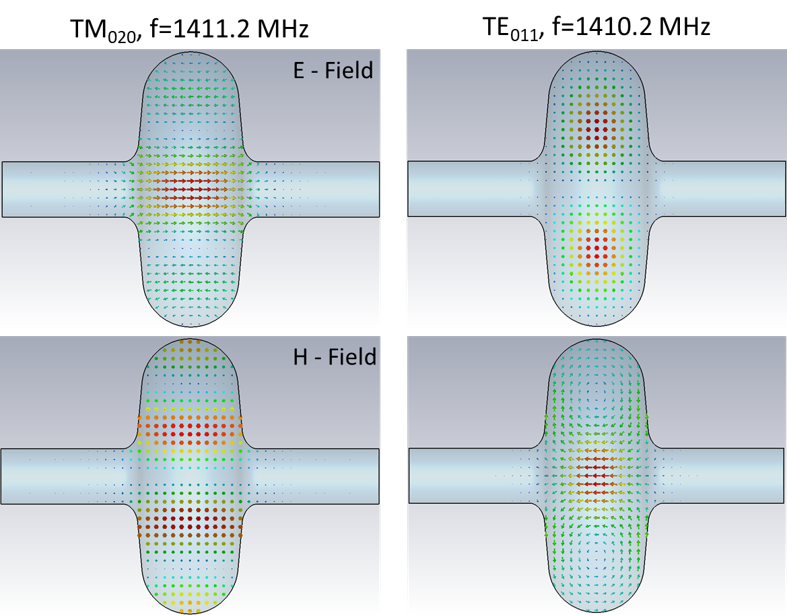}} \quad
	\caption{Finite element simulations of the electric and magnetic fields of the two resonant modes of the cavity used for the two-mode axion search.}
	\label{fig:2modefields}
\end{figure}
The SRF cavity designed at FNAL for this search is based on the geometry of a high$- \beta$ \SI{650}{\mega\hertz} single cell Nb cavity. Modes $\mathrm{{TM}_{020}}$ and $\mathrm{{TE}_{011}}$ were identified as pump and signal mode for the search, and their electric and magnetic field profiles are shown in Fig. \ref{fig:2modefields}. The cavity RF design was thoroughly studied and modified to achieve a \SI{1}{\mega\hertz} splitting between mode $\mathrm{{TM}_{020}}$ and $\mathrm{{TE}_{011}}$. In order to maximize coupling to the $\mathrm{{TE}_{011}}$ mode with minimal coupling to the $\mathrm{{TM}_{020}}$, the couplers for this mode were positioned symmetrically on the cavity wall in proximity to the node of the $\mathrm{{TM}_{020}}$ mode. Fig. \ref{fig:2modegeom} shows the final mechanical design of the 2-mode axion dark matter cavity. Loops antennas, shown in Fig. \ref{fig:2modeantenna}, will be used to couple to the $\mathrm{{TE}_{011}}$ magnetic field. Couplers for the $\mathrm{{TM}_{020}}$ will be positioned on the cavity beamtube axis.
The cavity will be equipped with a mechanical tuner to control the frequency splitting between pump and signal mode, enabling to scan over a range of $\approx \SI{2}{\mega\hertz}$ on the axion mass.

\subsection{Three-Mode Axion Search}
The search proposed by Bogorad \textit{et al.} \cite{bogorad2019probing} and further analyzed in Kahn \textit{et al.} \cite{10.1117/12.2616734} is similar to a light shining through the wall experiment (LSW) except that both production and detection of the axion is achieved in the same cavity. This search requires the use of two pump modes, one a TE, one a TM mode, excited simultaneously in the cavity volume. The axion would couple to specific linear combinations of the two pump modes, including $2 \omega_1 - \omega_2$. Interestingly, this experiment may not only allow to search for axions, but also to measure Euler-Heisenberg (EH) \cite{Heisenberg:1935qt,Schwinger:1951nm} induced nonlinearities. In order to detect either effects, the axion or the EH light-by-light scattering, the cavity should be designed so that $2 \omega_1 - \omega_2$ coincides with a third resonant frequency $\omega_3$ in order to enable the converted photons to build up in the cavity volume and to be detected on the transmitted signal. This experiment may enable the discovery of a new beyond the standard model particle or to help provide insight into nonlinear currents in superconductors.

\begin{figure}
    \centering
    {\includegraphics[width=1\columnwidth]{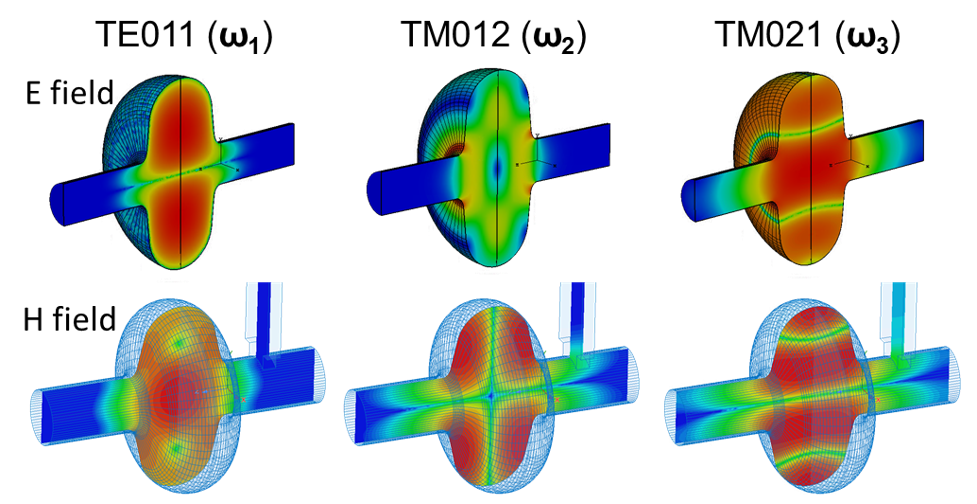}} \quad
	\caption{Finite element simulations of the electric and magnetic fields of the three resonant modes of the cavity used for the three mode axion search.}
	\label{fig:3modefield}
\end{figure}

\begin{figure}
    \centering
    {\includegraphics[width=0.8\columnwidth]{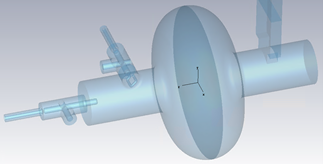}} \quad
	\caption{RF geometry for the three-mode axion search. The design of this cavity is still under study.}
	\label{fig:3modegeom}
\end{figure}
The RF design of the three mode axion cavity is still under study. Nevertheless, Fig. \ref{fig:3modefield} shows the electric and magnetic fields of the three resonant modes that will be used as pumps and signal frequencies for the three-mode axion cavity. Fig. \ref{fig:3modegeom} shows the current status of the RF design of the single cell cavity. The RF design of the three-mode axion cavity offers many complications as all three resonant frequencies fall between few hundreds of MHz from each other. Since the detection scheme complies on the use of a low noise RF amplifier operating at cryogenic temperature, it is crucial to filter room temperature noise at the frequency of the signal mode from entering the cavity from the two input couplers. In addition, it is necessary to minimize crosstalk and power leaks between the two input couplers, and to filter the pump modes from the transmitted signal to avoid RF intermodulation in the readout instrumentation that may generate a signal at the expected axion frequency. Multiple design solutions are being explored in order to address each of these challenges.

\section{\label{sec:Conclusions}Conclusions}
This work presents the current status of some of the axion and axion dark matter searches that are being prepared by the Physics and Sensing thrust at SQMS. The searches here discussed will be based on SRF Nb cavities and will use multiple resonant modes to produce the axions (in some cases) and to detect the axion-coupled photons. The design of Berlin \textit{et al.} \cite{berlin2020axion} two-mode axion cavity is completed, the design of Bogorad \textit{et al.} \cite{bogorad2019probing} is currently under study and close to be finalized, while work on Gao and Harnik \cite{gao2021axion} LSW search will soon start.

In preparation to the design of the SRF cavities for axion searches, we conducted a feasibility study to identify possible sources of nonlinearities in the cavity volume that could limit the search sensitivity. In the single-mode measurements we excluded energy transfer from an excited mode with $P_{in}=\SI{4}{\watt}$ to other resonant frequencies $\approx \SI{1}{\giga\hertz}$ away, with background limited by room temperature thermal noise measured with BW $=\SI{1}{\hertz}$. We also excluded energy leakage from an excited mode to resonant frequencies $1-$\SI{30}{\mega\hertz} away when $P_{cavity} \approx$ -3dBm, with background in the $\mathrm{{TM}_{010}}$ frequency range varying between -150 and -160 dBm. During the multi-mode measurements peaks at the linear combinations of the cavity pump modes were measured, however, we could not conclude in a definitive way that the signals were generated in the cavity volume. More studies will follow on this topic. Input from this study will be valuable in the design of the entire experimental system that will be used to run the axion searches and already provided valuable input regarding possible noise sources that may significantly limit search reach.

\begin{acknowledgments}
This material is based upon work supported by the U.S. Department of Energy, Office of Science, National Quantum Information Science Research Centers, Superconducting Quantum Materials and Systems Center (SQMS) under contract number DE-AC02-07CH11359.
\end{acknowledgments}

\bibliography{AxionCavities}

\end{document}